\documentclass[12pt,preprint]{aastex}

\slugcomment{}

\shorttitle{Variability in WISEP J173835.52$+$273258.9}
\shortauthors{Leggett et al.}

\begin{document}


\title{Observed Variability at 1\,$\mu$m and 4\,$\mu$m 
in the\\ Y0 Brown Dwarf WISEP J173835.52$+$273258.9}


\author{S. K. Leggett\altaffilmark{1}}
\email{sleggett@gemini.edu}
\author{Michael C. Cushing \altaffilmark{2}}
\author{Kevin K. Hardegree-Ullman \altaffilmark{2}}
\author{Jesica L. Trucks \altaffilmark{2}}
\author{M. S. Marley\altaffilmark{3}}
\author{Caroline V. Morley\altaffilmark{4}}
\author{D. Saumon\altaffilmark{5}}
\author{S. J. Carey \altaffilmark{6}}
\author{J. J. Fortney \altaffilmark{4}}
\author{C. R. Gelino\altaffilmark{7,8}}
\author{J. E. Gizis \altaffilmark{9}}
\author{J. D. Kirkpatrick\altaffilmark{7}}
\author{G. N. Mace \altaffilmark{10}}

\altaffiltext{1}{Gemini Observatory, Northern Operations Center, 670
  N. A'ohoku Place, Hilo, HI 96720, USA} 
\altaffiltext{2}{The University of Toledo, 2801 West Bancroft Street, Mailstop 111, Toledo, OH 43606, USA}
\altaffiltext{3}{NASA Ames Research Center, Mail Stop 245-3, Moffett Field, CA 94035, USA}
\altaffiltext{4}{Department of Astronomy and Astrophysics, University of California, Santa Cruz, CA 95064, USA}
\altaffiltext{5}{Los Alamos National Laboratory, PO Box 1663, MS F663, Los Alamos, NM 87545, USA}
\altaffiltext{6}{Spitzer Science Center, CalTech, Pasadena, CA 91125, USA}
\altaffiltext{7}{IPAC, CalTech, Pasadena, CA 91125, USA}
\altaffiltext{8}{NASA Exoplanet Science Institute, CalTech, Pasadena, CA 91125, USA}
\altaffiltext{9}{University of Delaware, Newark, DE 19716, USA}
\altaffiltext{10}{University of Texas, Austin, TX 78712, USA}

\begin{abstract}
We have monitored photometrically the Y0 brown dwarf  WISEP J173835.52$+$273258.9 (W1738) at both near- and mid-infrared wavelengths. This $\lesssim 1$ Gyr-old 400~K dwarf is at a distance of 8~pc and has a mass around 5 M$_{\rm Jupiter}$. We observed W1738 using two near-infrared filters at $\lambda \approx 1\,\mu$m, $Y$ and $J$, on Gemini observatory, and two mid-infrared filters at $\lambda \approx 4\,\mu$m, [3.6] and [4.5], on the {\em Spitzer} observatory. Twenty-four hours were spent on the source by  {\em Spitzer} on each of June 30 and October 30 2013 UT. Between these observations, around 5 hours   were spent on the source by Gemini on each of July 17 and August 23 2013 UT.  The mid-infrared light curves show significant evolution between the two observations separated by four months.  We find that a double sinusoid can be fit to the [4.5] data, where one sinusoid has a period of $6.0 \pm 0.1$ hours and the other a period of $3.0 \pm 0.1$ hours. The near-infrared observations suggest variability with a $\sim 3.0$ hour period, although only at a $\lesssim 2 \sigma$ confidence level. 
We interpret our results as showing that the Y dwarf has a $6.0 \pm 0.1$ hour rotation period, with one or more large-scale surface features being the source of variability.    The peak-to-peak amplitude of the light curve at [4.5] is 3\%. 
The amplitude of the near-infrared variability, if real, may be as high as 5 to 30\%. Intriguingly, this size of variability and the wavelength dependence can be reproduced by
atmospheric models that include patchy KCl and Na$_2$S clouds and associated small changes in surface temperature.  
The small number of large features, and the timescale for evolution of the features, is very similar to what is seen in the atmospheres of the solar system gas giants.

\end{abstract}

\keywords{stars: brown dwarfs, stars: atmospheres, stars: individual (WISEP J173835.52$+$273258.9)}

\section{Introduction}

There are now more than twenty brown dwarfs known in the solar neighborhood that have effective temperatures ($T_{\rm eff}$) lower than 500~K  (Cushing et al. 2011, 2014; Kirkpatrick et al. 2012; Liu et al. 2012; Luhman 2014; Luhman, Burgasser \& Bochanski 2011, Pinfield et al. 2014; Schneider et al. 2015; Tinney et al. 2012). These have been classified as Y dwarfs   (Cushing et al. 2011, Kirkpatrick et al. 2012). Evolutionary models   show that for $300 \leq T_{\rm eff}$~K $\leq 500$ and 0.2 $\leq$ age~Gyr $\leq$ 10 (as appropriate for the solar neighborhood) the mass range is 2 -- 30 Jupiter masses (Saumon \& Marley 2008). Hence this population of isolated brown dwarfs has a mass that is very planet-like.

Our group has an ongoing program measuring the photometric variability of Y dwarfs. 
For warmer brown dwarfs variability is usually associated with inhomogeneous or variable cloud structure in the atmosphere (e.g. Radigan et al. 2012). 
For  Y0 and Y1 dwarfs with $T_{\rm eff} \approx 400$~K the atmospheres are generally cloud-free, because most of the atmosphere is too cold for chloride or sulphide clouds, and too warm  for  water or ammonia clouds
(Burrows et al. 2003; Morley et al. 2012, 2014); in fact
cloud-free models can reproduce Y dwarf observations (Leggett et al. 2015, 2016).
Nevertheless variability may be seen at wavelengths where flux is emitted from very high and cold or low and warm layers where condensates can be present (Morley et al. 2012), or variability may be seen due to temperature variations across the brown dwarf surface (Showman \& Kaspi 2013). 

In our first paper (Cushing et al. 2016, hereafter Cu16) we show that the Y0.5(pec) brown dwarf WISEPC J140518.40$+$553421.5  (W1405, Cushing et al. 2011) is variable at  mid-infrared wavelengths. 
W1405 was observed with {\em Spitzer} using the IRAC camera (Fazio et al. 2004)
in the [3.6] and [4.5] filters. Variability was evident at [4.5] in the first epoch and at both [3.6] and [4.5] in the second epoch. The  second-epoch light curves have a period of about 8.5 hr, and semi-amplitudes of 3.5\%.
In the current paper we present the detection of variability at mid-infrared wavelengths in another Y0, 
WISEP J173835.52$+$273258.9 (W1738, Cushing et al. 2011). We also present the tentative  detection 
of variability at near-infrared wavelengths, at the $\lesssim 2 \sigma$ confidence level. 
We extend to lower limits the work of Rajan et al. (2015) who  excluded any $J$-band variability larger than 20\% for this brown dwarf. 

Leggett et al. (2016) compares near-infrared spectra and photometry for W1738 to recent models which include chemical disequilibrium driven by vertical transport (Tremblin et al. 2015). It was necessary to include mixing in order to reproduce the observations, and a  cloud-free  solar metallicity model with   $T_{\rm eff} = 425 \pm 25$K and log $g = 4.0 \pm 0.25$ fit the data best. This temperature and relatively low gravity imply that W1738 is a 
3 -- 9 Jupiter mass object with an age of 0.15 -- 1 Gyr. Table 1 lists properties of W1738.

In \S 2 we present new observations of W1738 obtained with {\em Spitzer} and IRAC, and Gemini Observatory and its near-infrared imager NIRI (Hodapp et al. 2003). We obtained two epochs of  [3.6] and [4.5] data, separated by four months, as well as two epochs of near-infrared $Y$ and $J$ data, obtained between the  {\em Spitzer} observations and separated by one month. 
\S 3 presents our analysis of the data, which we discuss in \S 4. Our conclusions are given in \S 5.

\section{Observations}

\subsection{Gemini NIRI Observations}

W1738 was observed as part of the Gemini North program GN-2013A-Q-21. The brown dwarf was imaged in the  $Y$ and $J$ filters using NIRI.  The NIRI $Y$ filter differs slightly from the MKO standard, where $Y_{\rm NIRI} - Y_{\rm MKO} = 0.17\pm 0.03$ magnitudes for late type T and early type Y dwarfs (Liu et al. 2012).

Long-duration observations were obtained on UT 2013 July 17 and 2013 August 23 in photometric conditions with typical seeing $0\farcs8$. Integration times for both filters were 60\,s, and offsets of about 12$\arcsec$ were used in dither patterns that were moved slightly on the detector through the observation, to minimise the impact of bad pixels and reduce any flat-fielding artifacts. The filters were alternated so that a five-position dither in $Y$ was executed, followed by a nine-position dither in $J$, followed by another five-position dither in $Y$, etc. 

On 2013 July 17 the observation ran from UT 07:22:57 to 12:18:09, for a duration of 4.92 hours. Seventeen five-position dithers in $Y$ and sixteen nine-position 
dithers in $J$ were obtained, for an on-source time of 1.4 hours in $Y$ and 2.4 hours in $J$. On 2013 August 23 the observation ran from UT 5:41:48 to 10:19:56, 
for a duration of 4.64 hours. Sixteen five-position dithers in $Y$ and fifteen nine-position dithers in $J$ were obtained, for an on-source time of 1.33 hours in 
$Y$ and 2.25 hours in $J$. Photometric standard FS 27 was observed before the W1738 observation, and FS 35 was observed after, on both nights. The data were 
reduced in the standard way, using dark and flatfield images obtained with the on-telescope calibration unit, and Gemini IRAF routines.

The NIRI detector suffers from first-frame pattern noise, which occurs in the first frame following any filter change. Because of this we discarded all of the first frames in each set, and used sets of four $Y$ images and sets of eight $J$ images to first generate sky frames and then coadded images. Aperture photomery with annular skies and an aperture diameter of 2$\farcs$4 was performed.  

Six stars (or point-like objects) with $Y$ and $J$ magnitudes between 17 and 18 provided the photometric baseline reference. We used five point sources with  $Y$ and $J$ magnitudes between 18.9 and 20.7 to further explore the precision of the photometry, in order to determine whether or not the similarly-bright W1738 is variable in these filters. The eleven sources were selected for brightness, and for being present in every coadded image with a clean annular sky region.
The uncertainty in the photometry obtained from the four-minute $Y$ and eight-minute $J$ coadded images was 1 -- 2\% for the brighter reference stars in both filters, and $\sim$5\% in $Y$ and $\sim$3\% in $J$ for W1738 and the five fainter point sources. Figure 1 shows examples of a
coadded four-minute $Y$ and eight-minute $J$ image, with W1738 and the eleven other sources identified. Table 2 gives the nightly average of the photometry for W1738  and the other sources. The last two coadded images for each filter on the night of 2013 August 23 suffered from poor seeing, and while the data were used for relative photometry, the absolute values in Table 2 exclude these two datapoints.

We compared the photometry of W1738 and the five fainter sources to the six brighter reference stars, and also compared each bright reference star to the other five reference stars as a check on data quality. Figures 2 and 3 show the resulting light curves for each night. These are discussed below in \S 3.

\subsection{{\em Spitzer} IRAC Observations}

W1738 was observed in both the [3.6] and [4.5] filters of the IRAC camera on {\em Spitzer}, with the filters observed consecutively. For each filter 12.0 hour 
long observations were obtained, and the 24.0 hour long observation with the filter pair was repeated four months later. The first epoch ran from UT 2013 June 29 
19:22:39.36 to 2013 June 30 07:23:48.48 for [3.6], and 2013 June 30 07:30:00.00 to 19:31:26.40 for [4.5]. The second epoch ran from UT 2013 October 29 
18:29:22.56 to 2013 October 30 06:30:31.68 for [3.6], and 2013 October 30 06:36:25.92 to 18:37:52.32 for [4.5]. The data were obtained as part of program 90015 
during campaigns 35100 and 35900. The data were obtained and reduced in the same way as that for W1405, as decribed in Cu16. Briefly, the 
100-second images were obtained in ``staring" mode for each filter. Photometric analysis starts with the Basic Calibrated Data frames, which are converted from 
units of MJy sr$^{-1}$ to electrons. 
Aperture photometry with a radius of 3 pixels and a background annulus is then obtained using custom 
Interactive Data Language (IDL) code. For W1738 the field was more crowded than that of W1405, and extra care had to be taken with placement of 
the sky annuli. Data points that were extreme outliers, exceeding the median by 50 times the median absolute deviation, were removed.
Figure 4 shows the light curves obtained. The apparently bright data points are due to hot pixels which are caused by cosmic rays.

The average  brightness of the target did not change significantly between June and October 2013. We derive from the processed mosaics downloaded from the {\em Spitzer} archive [3.6] $=  16.89 \pm  0.08$ (where the uncertainty is dominated by the uncertainty in the sky value) and [4.5] $= 14.46 \pm 0.01$.

\section{Results: Variability  of W1738}

Figure 4 indicates that W1738 is variable in the [4.5] bandpass with an approximately six-hour period. We ran a simple  Fourier Transform analysis for each of the two [4.5] data sets, not taking into account the truncation of the time series. Figure 5 plots the result --- a  strong peak is found at a period of 6.0 hours for the 2013 June 30 data, and less significant 
peaks at 3.0 and 6.0 hours for the 2013 October 30 dataset. We also ran a Lomb-Scargle periodogram analysis on both the [3.6] and [4.5] data, on the two epochs. A significant period was only found for the 2013 June 30 [4.5] data, with a broad 
peak in power between 5 and 7 hours.

The near-infrared light curves shown in Figures 2 and 3 suggest that W1738 may also vary at near-infrared wavelengths. 
In Figure 2,  both $Y$ and $J$ brighten at around 9 and 12 hours. Trends are less clear in Figure 3, but there 
appears to be another three hour cycle present for both filters, with a minimum around 7.9 hours and maxima at 6.4 and 9.4 hours.  Table 3 compares the peak-to-peak variation (range) and standard deviation of the curves obtained for W1738 and the five faint point sources (numbered 2, 4, 8, 9, and 11  in Figure 1).  The stars are listed from faintest to brightest in the Table. As expected, the range and standard deviation generally increase with decreasing brightness. 
Figures 2 and 3 show that the dispersion for W1738 is similar to that of the comparison star that is fainter, and larger than that of the similarly bright comparison star. The dispersion is largely driven by two or three  data points however and the Y dwarf should be monitored for a longer period of time to confirm the presence of any variability. 
The standard deviation of the light curves for the five reference stars is on average $1.2 \times$ the error, while that for W1738 is  $1.9 \times$ the error. The fact that the three hour cycle is also seen in the mid-infrared data, and that models calculate a $1\, \mu$m variability amplitude $\sim 10 \times$ the $4\, \mu$m variability amplitude (\S 4.2), indicates that the near-infrared variability may be real. We interpret the data as implying that W1738 is variable at  
$1\, \mu$m 
at the  $\lesssim 2 \sigma$ confidence level.

Based on the Fourier transform results and our visual inspection of the light curves, three- and six-hour cycles are present in the W1738 data.  
We therefore fit the [4.5] light curves assuming a double sinusoid model, where the second sinusoid has a period half the first. The amplitudes and phases of the two 
sinusoids can vary freely. A possible physical explanation of this double sinusoid, where the second period is half the first, is presented in \S 4.3. The 
model is: $$F(t) = A_1\sin\left(\frac{2\pi}{P}t+\phi_1\right) + A_2\sin\left(2\frac{2\pi}{P}t+\phi_2\right) + C$$ where $A$ is semi-amplitude in \%, $P$ is period in 
hours, $t$ is time in hours, $\phi$ is phase in radians and $C$ is a constant. 

The fitting procedure is described in detail in Cu16.
Briefly, we assume the uncertainties are gaussian and account for the bad data points following Hogg, Bovy \& Lang (2010) whereby we assume that they are generated from a normal distribution with a mean  $y_{\rm bad}$ and a variance of $\sigma_{\rm bad}$.  The joint posterior distribution of the parameters $A_1$, $A_2$, $\phi_1$, $\phi_2$, $P$, $C$, $\sigma_{\rm good}$, ${\cal P}_{bad}$ (the probability that a given data point is bad) is sampled using a Markov Chain Monte Carlo method. Distributions for each model parameter are computed by marginalizing over the other parameters.
 
The one and two dimensional projections of the posterior probability distributions of the model parameters for the 2013 June 30 [4.5] sequence is shown in Figure 
6. Figures 7 and 8 show the resulting best fits to the [4.5] data for the two epochs, and the residuals between fit and data.  As was also indicated by the 
initial Fourier transform, the data from 2013 June 30 is dominated by the longer-period sinusoid, while the 2013 October 30 data consists of two sinusoids with 
almost equal amplitude.  A good fit could not be achieved with the noisy [3.6] data, however extrapolating the sinusoids to earlier times shows 
that the [3.6] data are not inconsistent with variability at the same amplitude and phase as the [4.5] data (Figure 4). Table 4 gives semi-amplitude, period and 
phase for each of the two sinusoids on each epoch, using the [4.5] data only.

\section{Discussion}

\subsection{Observed Variability in Brown Dwarfs and Giant Planets}

There are no other published studies of variability at both near-infrared and mid-infrared wavelengths for Y dwarfs at the time of writing. Nine T dwarfs do have 
such data published and these results are summarized in Table 5. Variability is found in five of the nine T dwarfs.
Generally the periods are found to be the same at both near- and mid-infrared wavelengths, and the near-infrared amplitudes are similar to or larger than the mid-infrared amplitude
(although the near- and mid-infrared data are likely to not have been taken at the same time, and amplitudes are likely to vary with time).

The variability measurements presented here for W1738 are in general agreement with results from other brown dwarf studies.
Crossfield (2014) presents a multi-wavelength collation of variability amplitude and period, and projected rotational velocities, for late-type M, L and T dwarfs. The Crossfield sample of L and T dwarfs has variability periods of 1.5 -- 10 hours and  peak-to-peak amplitudes of 0.1 -- 120\%, although most have amplitudes of 1 -- 15 \%. The {\em Spitzer} study of 39 single L3 -- T8 dwarfs by Metchev et al. (2015) found that 19 varied; peak-to-peak amplitudes were 0.8 -- 4.6\% and periods were 1.6 -- 24 hours. 
Recently Zhou et al. (2016) have determined a rotation period of about 11 hours for the young planetary-mass L dwarf 2MASSWJ 1207334-393254b, based on near-infrared variability with  peak-to-peak amplitudes of $\sim$ 2\%.
In Cu16 we monitored the Y0.5(pec) W1405 with  {\em Spitzer} and found  variability with  peak-to-peak amplitude of 7\% and a period of 8.5 hours. 
Cu16 showed that the variability could be reproduced by a single bright spot model, with the light curve period equal to the rotational period, meaning that W1405  has the longest rotation period measured to date for spectral types later than T3.  

Variability seen in the surface features of the solar system gas giants are also a useful reference point for cold Y dwarfs.  Gelino \& Marley (2000) used observations of Jupiter to show that, were it to be unresolved, 20\% variability would be observed at $4.8\, \mu$m with a period equal to the planet's rotation period of 9.9 hours. The variability in this case is predominantly due to Jupiter's Great Red Spot. 
Sromovsky et al. (2012) present observations of episodic bright and dark spots on Uranus. One and then two bright spots were seen on the planet's surface in 2011, drifting at very different rates, and evolving over a period of months. 
Simon et al. (2016) present a 49-day light curve for Neptune using the $Kepler$ Space Telescope. The data are compared to contemporaneous images taken with the Keck telescope at $1.65\, \mu$m and to {\em Hubble Space Telescope (HST)} visible imaging taken several months later. The authors find that a single large, long-lived, storm that is seen in their Keck images dominates the $Kepler$ and  $HST$ light curves. The periodicity of the long-term variability is consistent with the planet's rotation and surface wind speed at the latitude of the storm. The short-term variability  is interpreted as being due to smaller or fainter clouds.

\subsection{Models of Variability in Brown Dwarfs and Giant Planets}

Three-dimensional simulations of convection in brown dwarf and giant planet interiors by  Showman \& Kaspi (2013)
show that significant circulation is generated at both small and large scales.   Large-scale horizontal temperature variations of tens of K are produced, resulting in flux variations of a few percent on rotation timescales. Stratified turbulence can generate vortices and storms, and the circulation can support the formation of patchy clouds. These models
produce  vertical velocities consistent with the chemical mixing observed in brown dwarf atmospheres, plausible surface wind speeds, and time scales for the evolution of the light curve that are in agreement with observations.

It is likely that both thermal and cloud cover fluctuations are important sources of variability in giant planets and brown dwarfs (e.g. Morley et al. 2014, Robinson \& Marley 2014).
Although the $T_{\rm eff} \approx 400$ K atmospheres of Y dwarfs are expected to be essentially clear, Morley et al. (2012, their Figure 4) show that flux emitted from the $1\, \mu$m region originates deep enough in the atmosphere that it could be impacted by thin layers of KCl and Na$_2$S clouds. Morley et al. (2014) examine spectral variability that occurs if the clouds are patchy. Figure 1 of  Morley et al. (2014) shows the 0.7 -- $10\, \mu$m spectrum of a 400~K brown dwarf where one hemisphere has 30\% cloud cover and the other 70\%, and as a result also differ by 5~K or $> 1$\% of the $T_{\rm eff}$. The difference in the resulting spectra, which is the inferred peak-to-peak variability, is calculated to be $\sim$ 50\% at $1\,\mu$m ($Y$), 40\% at $1.2\,\mu$m ($J$) and 2\% at $4.5\, \mu$m. We ran a similar model with a covering fraction of 45\% on the Eastern hemisphere and 55\% on the Western hemisphere, which produced peak-to-peak variability of  7\% in $Y$, 5\% 
in $J$, 0.5\% in [3.6] and 0.8\% in [4.5]. Thus, atmospheres with patchy clouds and small variations in surface temperature can reproduce the size of variability we observe at [4.5], and have tentatively observed at $Y$ and $J$.

\subsection{Interpretation of the W1738 Variability}

Physical constraints on the rotation period of brown dwarfs can be helpful in determining the source of any observed variability. If the variation is caused by 
multiple surface features such as clouds or several discrete storms, then it can have a period smaller than the rotation period. If it is due to a single feature 
such as Jupiter's Red Spot or the single bright spot model used to reproduce the W1405 variability, then the period will be equal to the rotation period. The 
lower limit on the rotation period of any stable object can be estimated by assuming solid body rotation and constraining the surface velocity to be less than 
the escape velocity. Adopting a radius equal to that of Jupiter's (which is approximately true for most brown dwarfs, see Burrows et al. 1997), we find that the 
period $P > 2.1 / \sqrt M$ hours, where the mass of the brown dwarf is $M$ Jupiter masses. Marley \& Sengupta (2011, their Figure 2) more accurately derive a 
breakup velocity, and show that for brown dwarfs or gas planets older than 0.1 Gyr the lower limit on rotation period is 5 hours or 1.8 hours for a 1 or a 10 
Jupiter mass object, respectively. For W1738 with mass $\approx$ 5 Jupiter masses and age between 0.15 and 1.0 Gyr (Leggett et al. 2016), the rotational period 
must be greater than about 3 hours. We have found two dominant periodicities, three hours and six hours, in our data sets. The three hour 
component is very close to the break up speed
-- which is very unlikely -- suggesting that there are multiple features on the surface of the brown dwarf.

Metchev et al. (2015) found substantial power in periods approximately half of the best-fit period for three 
of the 19 variable L and T dwarfs in their sample.   Studies of pre-main-sequence stars have also found photometric variability with periods separated by a factor of two (Tackett, Herbst \& Williams 2003; Herbst 1989).
The pre-main-sequence observations could be reproduced by a model where the 
rotation period was equal to the longer of the two periods, the
star was viewed close to equator-on, and there were two similar spots, one on each hemisphere. The model showed that the shape and amplitude of the light curve evolved as the spots drifted in longitude (Herbst 1989).  
We attempted to fit a model with two equally bright spots to the W1738 data, where the spots are circular and  have their own longitude, latitude and size. The first epoch, where there is a dominant sinusoid, could be reasonably well reproduced by a two spot model, but not the second epoch, where there are two sinusoids of different frequency of almost equal amplitude. We did not explore models with more than two spots.

We adopt the rotation period for W1738 to be the longer of the two periods observed, $6.0 \pm 0.1$ hours, and interpret the double-sinusoid variability as evidence of there being one or more large features present in its atmosphere. The features evolve over a period of months, such that in June 2013 the variability is dominated by a single spot or storm, and by October 2013 smaller features have developed giving rise to the shorter period variability observed. This configuration of a dominant large system which is long-lived, and multiple smaller surface features, 
is similar to what was observed for Neptune by Simon et al. (2016, \S 4.1). We note that  for W1738 the amplitude of the variability, and the suggested wavelength dependence of the amplitude, is well reproduced by models with patchy thin clouds (\S 4.2). The models have one hemisphere slightly more than half-covered and the other slightly less than half-covered with clouds. A hemisphere half-covered in clouds could also be described as a hemisphere with a large spot, depending on the wavelength involved, and the height of the clouds.

\section{Conclusion}

We obtained 12 hours of continuous {\em Spitzer} data on the Y0 brown dwarf W1738 at [3.6], followed by another 12 hours at [4.5]. Two sets of data were obtained 
four months apart, on June 30 and October 30 2013. We also obtained interspersed Gemini $Y$ and $J$ data on W1738, with about 1.4 hours on-source at $Y$ and 2.3 
hours on-source at $J$, on two occasions separated by about one month, July 17 and August 23 2013.
 
Fourier and Lomb-Scargle analyses of the mid-infrared [4.5] data suggested the presence of three and six hour periods in the data.  We use a probabilistic method to fit double sinusoids to the [4.5] data (the near-infrared data covers a short time span and the [3.6] data are 
too noisy). We constrain the second sinusoid to have half the period of the first, but allow amplitude and phase to vary. We find sinusoids with periods of 5.8 
$\pm$ 0.1 hours and 6.13 $\pm$ 0.08 hours, and half those values, reproduce the observed [4.5] light curves on the two epochs well. The amplitudes range from 0.3\% 
to 1.1\%, leading to peak-to-peak variability of 3\%. The shorter time-span near-infrared data was inspected visually only. The data suggest that W1738 is also variable in the near-infrared, although only at the $\lesssim 2 \sigma$ confidence level. If real, the implication is that this Y0 dwarf varied by 10 -- 30\% in $Y$ and 5 -- 15 \% in $J$, peak-to-peak, with a period of about three hours, at two epochs during 2013. 

The observations are consistent with W1738, a 5 Jupiter-mass $T_{\rm eff} \approx 400$ K Y-type brown dwarf, being seen nearly-equator on, having a rotation 
period of $6.0 \pm 0.1$ hours, and having one or more large surface features which give rise to the variability. The features evolve over timescales of months. The 
observed variability at $\lambda \sim 4\,\mu$m is likely due to thermal variations caused by atmospheric circulation, while the larger variations at $\lambda 
\sim 1\,\mu$m, if real, may be due to the presence of patchy clouds of KCl and Na$_2$S in the lower regions of the atmosphere.

\acknowledgments

Based in part on observations obtained at the Gemini Observatory, which is operated by the Association of Universities for
Research in Astronomy, Inc., under a cooperative agreement with the NSF on behalf of the Gemini partnership: the National Science Foundation (United States), the Science and Technology Facilities Council (United Kingdom), the National Research Council (Canada), CONICYT (Chile), the Australian Research Council (Australia),
    Minist\'{e}rio da Ci\^{e}ncia, Tecnologia e Inova\c{c}\~{a}o (Brazil)
    and Ministerio de Ciencia, Tecnolog\'{i}a e Innovaci\'{o}n Productiva
    (Argentina). 
Based in part on observations obtained with the {\em Spitzer Space Telescope}, which is operated by  the Jet Propulsion Laboratory, California Institute of Technology under a contract with NASA. 
S. L.'s research is supported by Gemini Observatory.  D.S.' work was supported in part by 
NASA grant
      NNH12AT89I from Astrophysics Theory. 

We thank the referee for comments which greatly improved the paper.

\clearpage

\begin{figure}
\plottwo{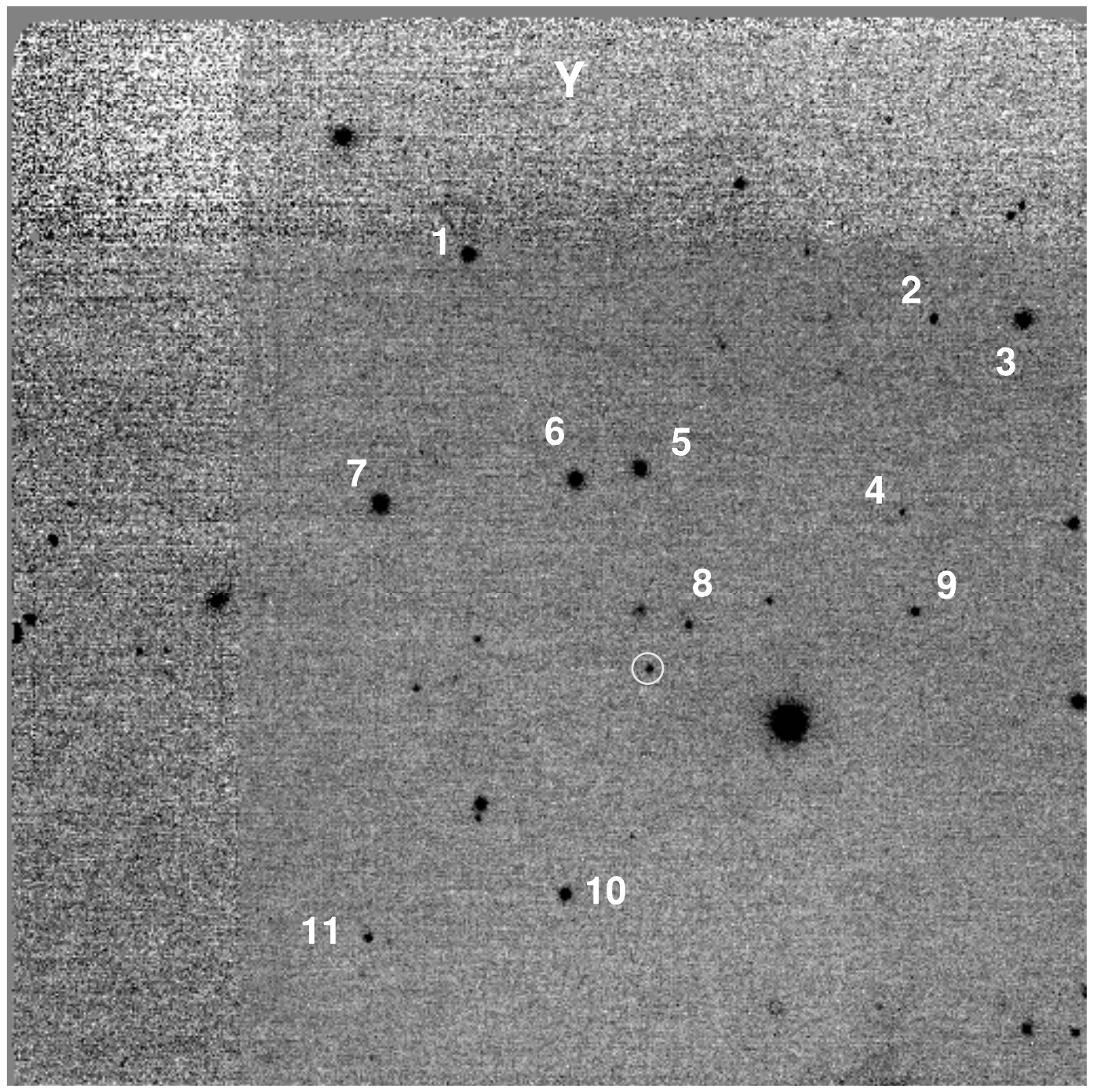}{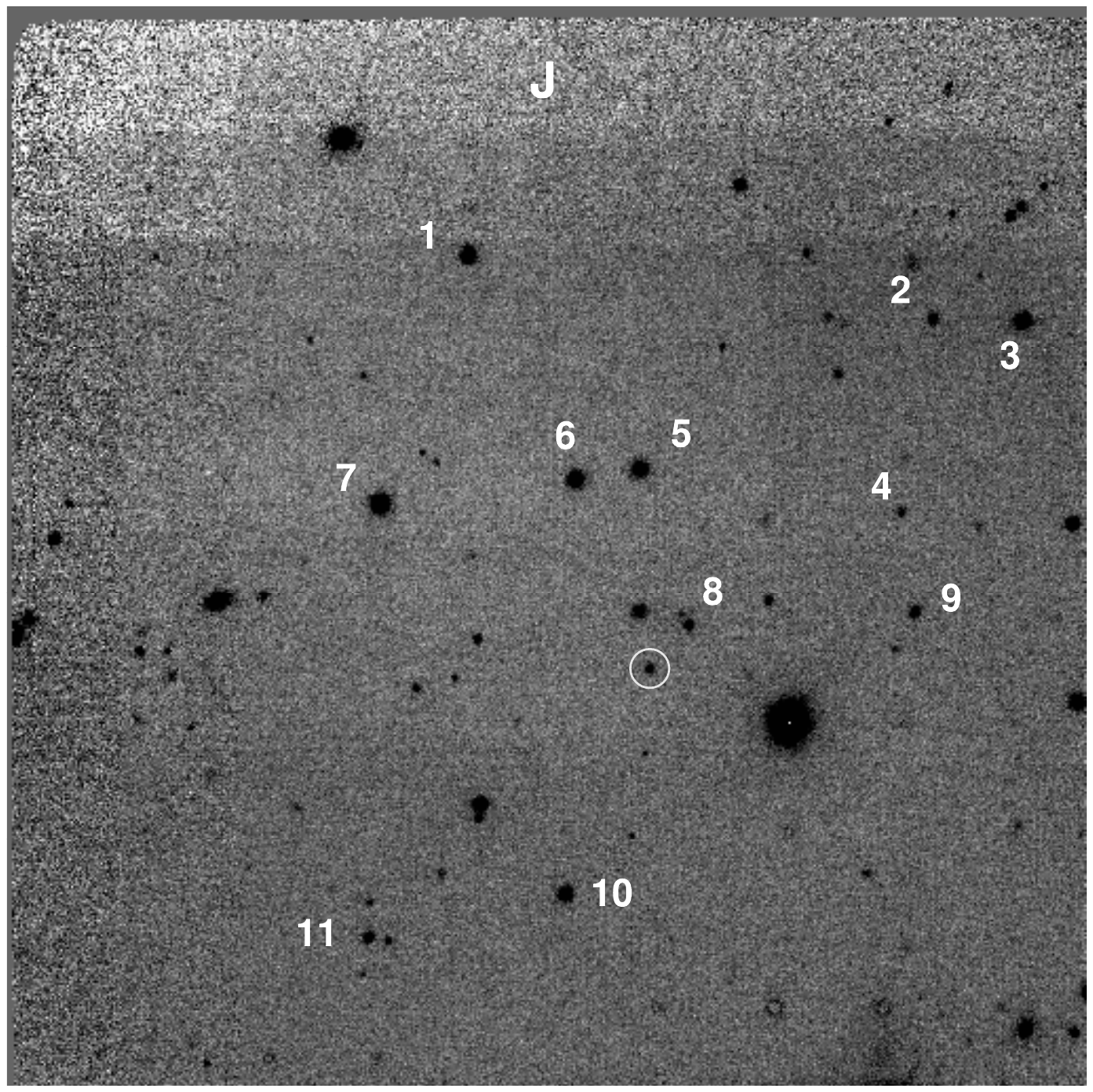}
\caption{NIRI images of W1738 in $Y$ (left) and $J$ (right). The field is 2 arcminutes on a side, with North up and East to the left. W1738 is circled, and the eleven photometric comparison stars are numbered.
\label{fig1}}
\end{figure}

\begin{figure}
    \includegraphics[angle=0,width=.8\textwidth]{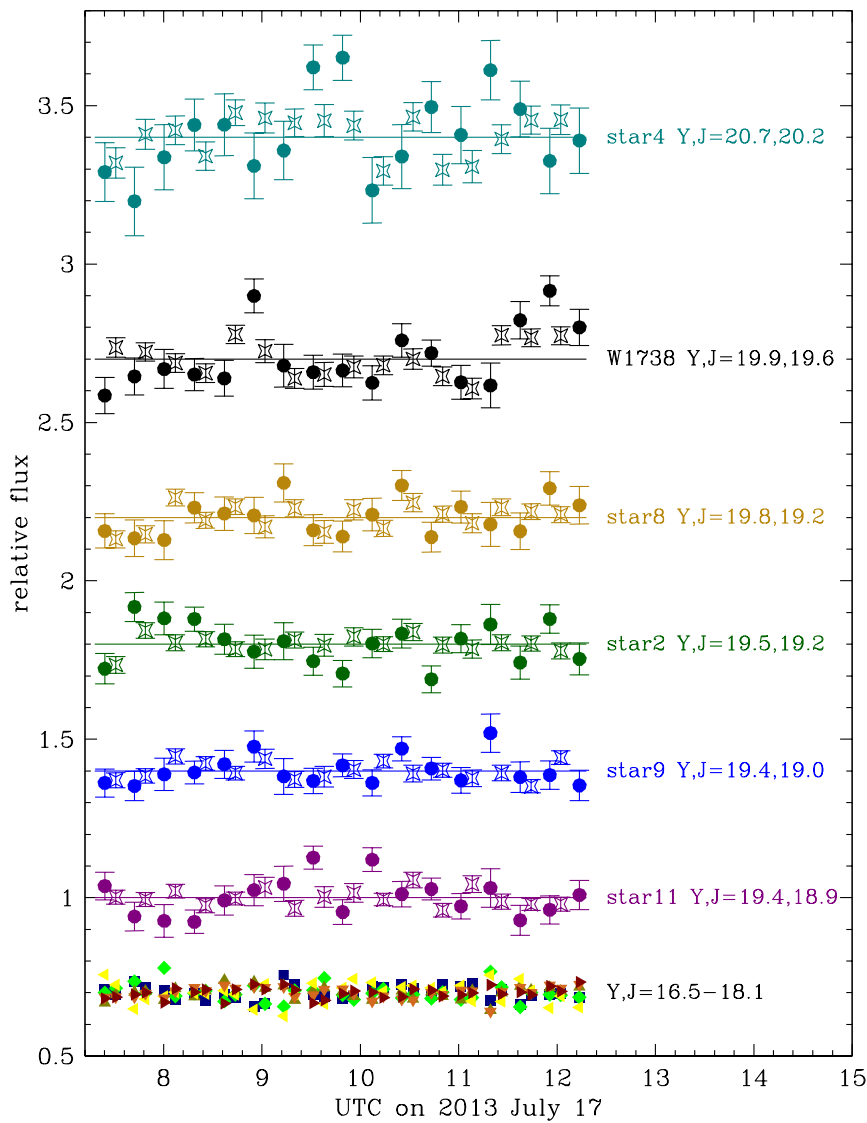} \caption{
Light curves for W1738 and five point sources identified in Figure 1, relative to the six brighter point sources in the field, on 2013 July 17. Solid circles represent $Y$ data and open stars $J$ data. 
The errors bars  are the square root of the sum 
of the squares of the individual measurement uncertainty and the standard deviation in the six reference measurements.  The grouping of six symbols along the bottom are the light curves for the six brighter sources in the field, where each has been compared to the other five.
\label{fig2}} \end{figure}

\begin{figure}
    \includegraphics[angle=0,width=.8\textwidth]{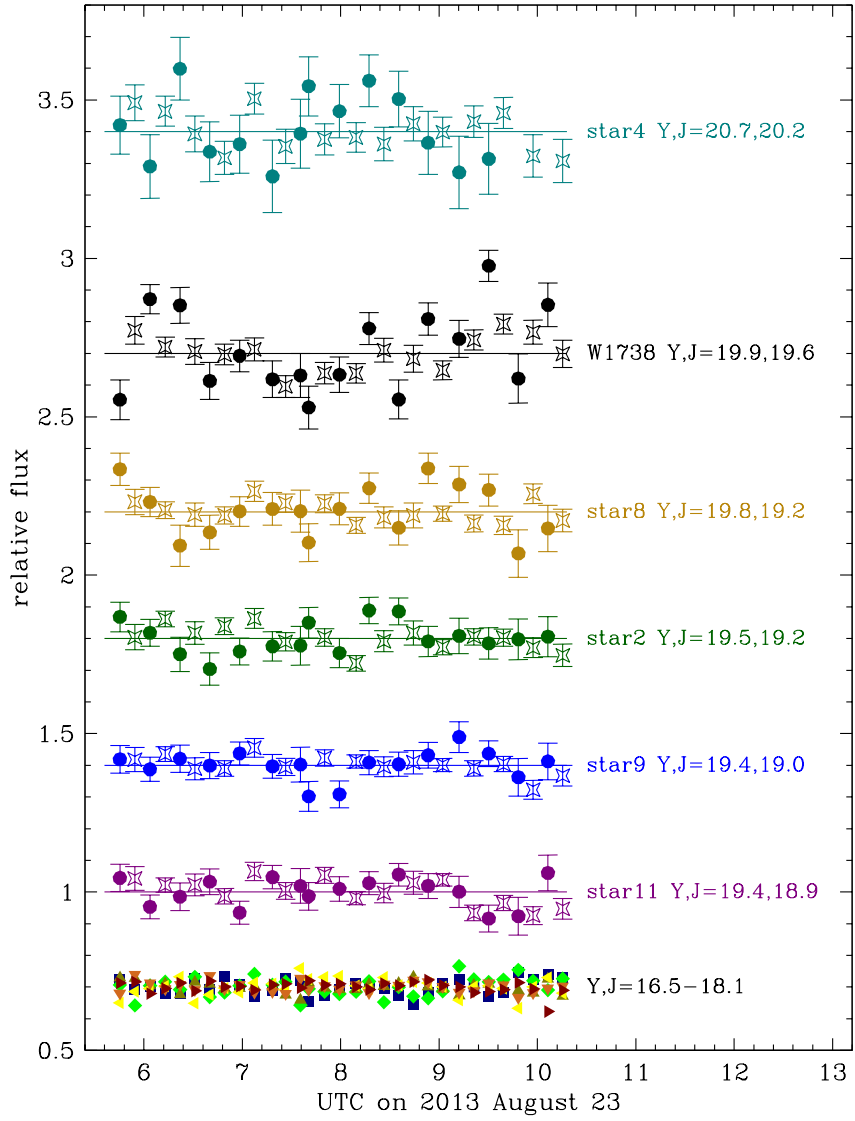}
\caption{The same as Figure 2, but for UT 2013 August 23.
\label{fig3}}
\end{figure}

\begin{figure}
    \includegraphics[angle=0,width=0.9\textwidth]{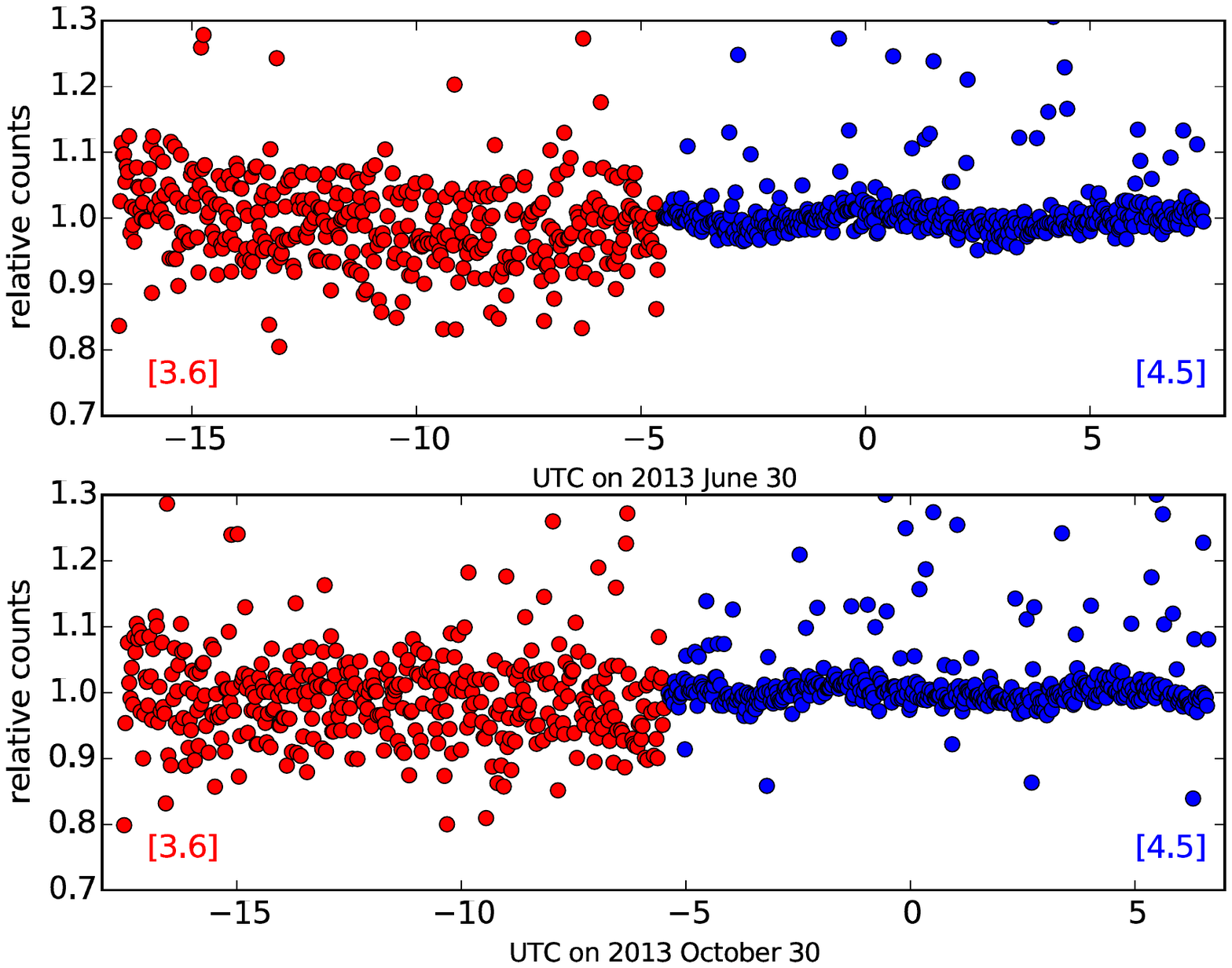}
\caption{Relative counts obtained with IRAC on {\em Spitzer} using the [3.6] and [4.5] filters, as a function of time on UT 2013 June 30 and October 30. The [3.6] signal is  fainter and  noisier. 
\label{fig4}}
\end{figure}

\begin{figure}
    \includegraphics[angle=0,width=0.8\textwidth]{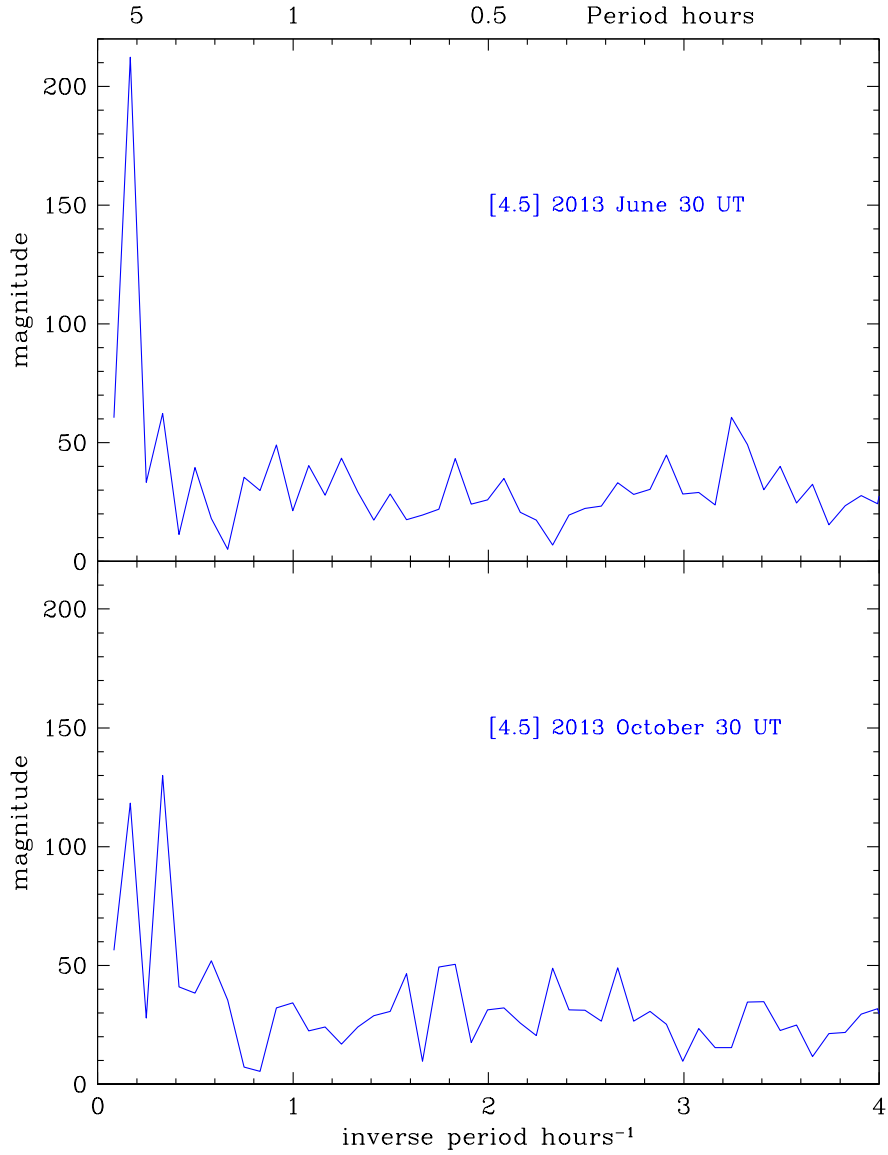} \caption{Results of a Discrete Fourier Transform analysis of the 12-hour data sets obtained with the 
[4.5] filter on 2013 June 30 and 2013 October 30, UT. Results are expressed as inverse period in hours, and magnitude. 
\label{fig5}} \end{figure}

\begin{figure}
    \includegraphics[angle=0,width=.99\textwidth]{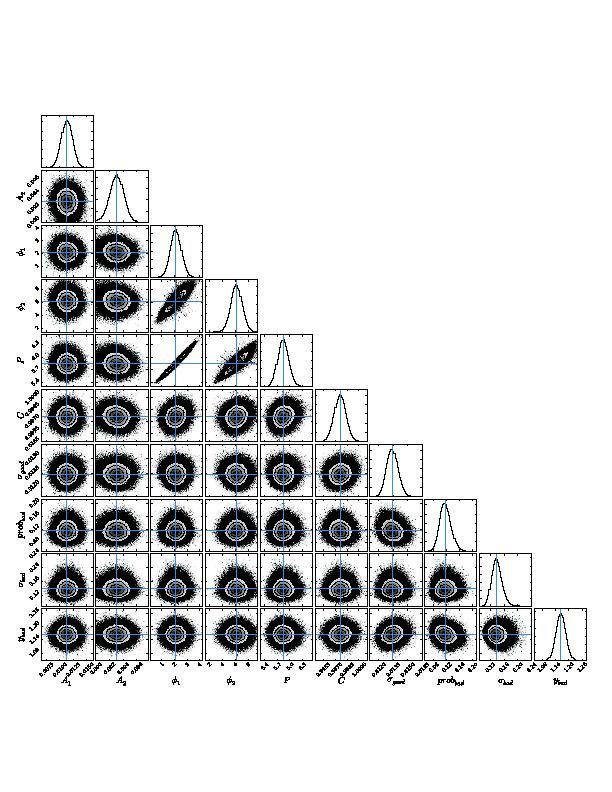}
\caption{Probability distribution for parameters describing a double sinusoid fit to the light curve measured on  UT 2013 June 30  in the IRAC [4.5] filter.
\label{fig6}}
\end{figure}

\begin{figure}
    \includegraphics[angle=-90,width=0.95\textwidth]{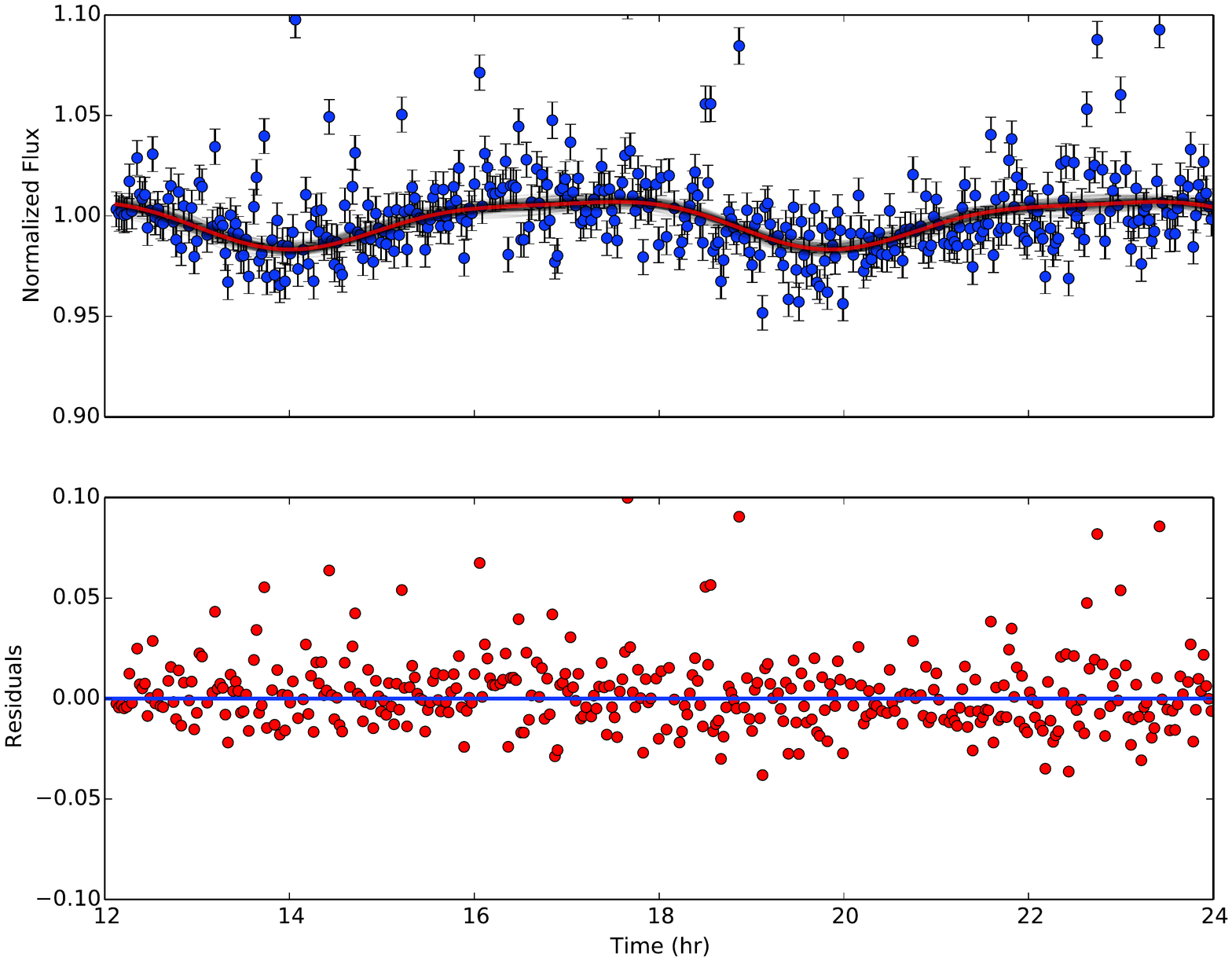}
\caption{The UT 2013 June 30 [4.5] light curve showing the best double sinusoid model fit (red) with parameters given in Table 4, and 100 randomly selected parameter sets from the MCMC chain (grey). Residuals shown in lower panel.
\label{fig7}}
\end{figure}

\begin{figure}
    \includegraphics[angle=-90,width=0.95\textwidth]{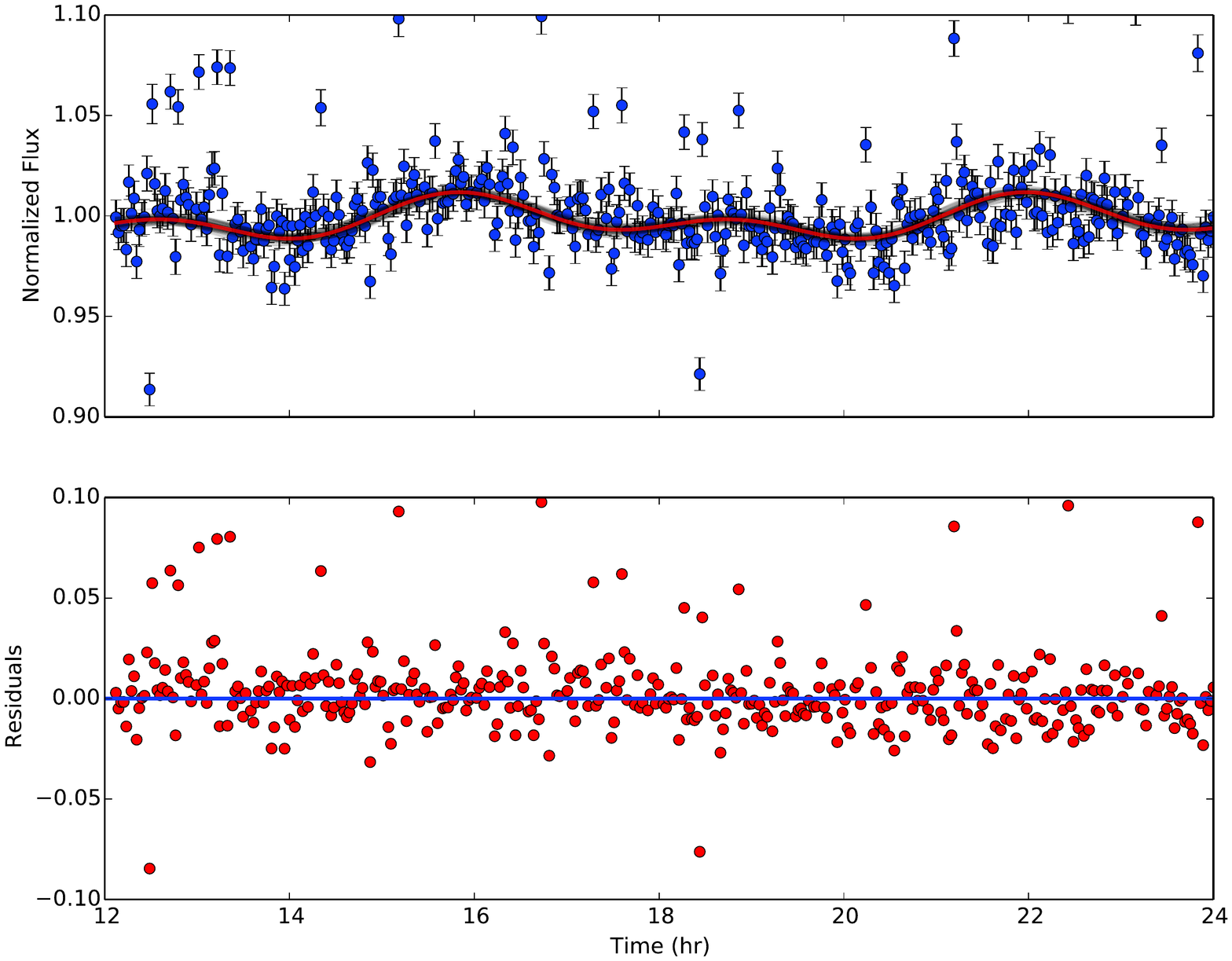}
\caption{The UT 2013 October 30 [4.5] light curve showing the best double sinusoid model fit (red) with parameters given in Table 4, and 100 randomly selected parameter sets from the MCMC chain (grey). Residuals shown in lower panel.
\label{fig8}}
\end{figure}

\clearpage

\begin{deluxetable}{lcc}
\tabletypesize{\footnotesize}
\tablewidth{0pt}
\tablecaption{Properties of WISEP J173835.52$+$273258.9}
\tablehead{ 
\colhead{Property}  & \colhead{Value}  & \colhead{Reference}  \\
}
\startdata
Spectral Type &  Y0 & Cu11\\
Distance pc & 7.8 $\pm$ 0.6 & Be14 \\
$V_{\rm tan}$ km s$^{-1}$ & 17   $\pm$ 1 & Be14 \\
Rotation period hr & 6.0 $\pm 0.1$ & this work \\
$T_{\rm eff}$ K & 425 $\pm$ 25 & Le16 \\ 
log $g$ cm s$^{-2}$&  4.0 $\pm 0.25$ & Le16 \\ 
Mixing coefficient $K_{\rm zz}\,$cm$^2\,$s$^{-1}$ & $10^6$ & Le16\\
Mass Jupiter & 3 -- 9 & Le16 \\ 
Age Gyr &  0.15 -- 1 & Le16 \\ 
$Y_{\rm MKO}$ mag &  19.74 $\pm 0.08$ & this work \\
$J_{\rm MKO}$ mag &   19.58 $\pm 0.04$  & this work \\
$H_{\rm MKO}$ mag &  20.24 $\pm 0.08$ & Le16 \\
$K_{\rm MKO}$ mag & 20.58 $\pm 0.10$ & Le13 \\
Ch.1(3.6~$\mu$m)$_{\rm IRAC}$ mag &  16.87 $\pm 0.03$ & Le13 \\
Ch.2(4.5~$\mu$m)$_{\rm IRAC}$ mag &  14.42 $\pm 0.03$ & Le13 \\
W1(3.4~$\mu$m)$_{WISE}$ mag &  17.71 $\pm 0.16$ &     AllWISE \\
W2(4.6~$\mu$m)$_{WISE}$ mag & 14.50 $\pm 0.04$  &   AllWISE \\
W3(12~$\mu$m)$_{WISE}$ mag & 12.45 $\pm 0.40$ &   AllWISE \\
$Y$ variability semi-amplitude \% & 5 -- 15 & this work, $\lesssim 2 \sigma$ confidence \\
$J$ variability semi-amplitude \% & 3 -- 8 & this work, $\lesssim 2 \sigma$ confidence \\
Ch.2(4.5~$\mu$m) variability semi-amplitude \% & 1.5 & this work \\
\enddata
\tablecomments{References are Beichman et al. 2014; Cushing et al. 2011;  Leggett et al. 2013, 2016.}
\end{deluxetable}

\clearpage

\begin{deluxetable}{lccccc}
\tabletypesize{\footnotesize}
\tablewidth{0pt}
\tablecaption{Nightly Averaged $Y$ and $J$ for WISEP J173835.52$+$273258.9 and Reference Stars}
\tablehead{ 
\colhead{Object} & \colhead{Nominal} & \multicolumn{2}{c}{2013 July 17 UT}   & \multicolumn{2}{c}{2013 August 23 UT} \\
\colhead{} & \colhead{RA Declination} & \colhead{NIRI Y ($\sigma$)} & \colhead{J ($\sigma$)}  & \colhead{NIRI Y ($\sigma$)} & \colhead{J ($\sigma$)} \\
\colhead{} & \colhead{hh:mm:ss.sss dd:mm:ss.s} & \multicolumn{4}{c}{magnitudes} \\
}
\startdata
W1738 & 17:38:35.598 27:32:58.16 & 19.88(0.10) & 19.55(0.06)  &  19.94(0.14) & 19.60(0.05)  \\
star1 & 17:38:37.100 27:33:43.8 & 17.63(0.03) & 17.21(0.02)  & 17.63(0.03) & 17.28(0.04) \\
star2 & 17:38:33.249 27:33:36.90 & 19.52(0.07) & 19.16(0.03) & 19.58(0.07) & 19.17(0.04) \\
star3 & 17:38:32.504 27:33:36.73 & 17.31(0.03) & 17.07(0.02)  &  17.35(0.03) & 17.13(0.02)  \\
star4 & 17:38:33.510 27:33:15.54 & 20.72(0.14) & 20.23(0.07) & 20.74(0.15) & 20.25(0.05) \\ 
star5 & 17:38:35.678 27:33:20.35 & 17.78(0.03) & 17.54(0.02) & 17.78(0.03) &  17.58(0.03) \\
star6 & 17:38:36.211 27:33:19.11 & 17.59(0.04) & 17.36(0.02) &  17.61(0.03) & 17.42(0.03) \\
star7\tablenotemark{a} & 17:38:37.829 27:33:16.46 & 16.76(0.03) & 16.47(0.02) & 16.77(0.02) & 16.53(0.03) \\
star8 & 17:38:35.273 27:33:02.99 & 19.79(0.06) & 19.23(0.04) & 19.83(0.08) & 19.25(0.04)\\
star9 & 17:38:33.401 27:33:04.49 & 19.38(0.04) & 18.99(0.03)& 19.38(0.06) & 19.00(0.03)\\
star10 & 17:38:36.298 27:32:33.2 & 18.12(0.04) & 17.66(0.02) & 18.10(0.03) &  17.71(0.03) \\
star11 & 17:38:37.926 27:32:28.44 & 19.35(0.07) & 18.86(0.03)& 19.37(0.06) & 18.93(0.05) \\
\enddata
\tablenotetext{a}{Star7 is 2MASS 17383786$+$2733160 with a 2MASS $J$ all-sky catalog magnitude of 16.598 $\pm$ 0.131. 
}
\end{deluxetable}

\clearpage

\begin{deluxetable}{lccccccccccccc}
\tabletypesize{\footnotesize}
\tablewidth{0pt}
\rotate
\tablecaption{Observed $1\, \mu$m Photometric Dispersion for  WISEP J173835.52$+$273258.9 and Five Faint Reference Stars}
\tablehead{ 
\colhead{Object} & \colhead{$Y$,$J$} &
\multicolumn{6}{c}{2013 July 17 UT}   & \multicolumn{6}{c}{2013 August 23 UT} \\
\colhead{}  & \colhead{Magnitude}  &
\multicolumn{3}{c}{Y \%} & \multicolumn{3}{c}{J \%}  & \multicolumn{3}{c}{Y \%} & \multicolumn{3}{c}{J \%} \\
\colhead{} & \colhead{} & 
\colhead{Range} & \colhead{$\sigma$} & \colhead{Error}  & \colhead{Range} & \colhead{$\sigma$} & \colhead{Error} &
\colhead{Range} & \colhead{$\sigma$} & \colhead{Error}  & \colhead{Range} & \colhead{$\sigma$} & \colhead{Error} \\
}
\startdata
star4 & 20.7,20.2 & 45.3 & 13.2 & 9.3 & 18.4 & 6.7 & 4.6 & 33.9 & 11.2 & 9.8 &  20.0 & 6.3 & 5.3 \\
W1738 & 19.9,19.6 & 33.1 & 10.0 & 5.6 & 17.1 & 5.5 & 3.1 & 44.7 & 13.5 & 5.9 & 19.7 & 5.6 & 3.6 \\
star8 & 19.8,19.2 & 18.0 & 5.9 & 5.4 & 13.2 & 3.9 & 2.8 & 26.8 & 8.3 & 5.6 & 10.8 & 3.4 & 3.1 \\
star2 & 19.5,19.2 & 22.8 & 6.8 & 4.8 &  10.9 & 2.6 & 2.6 & 18.6 & 5.1 & 5.0 & 14.2 & 3.8 & 3.0 \\
star9 & 19.4,19.0 & 16.6 & 4.7 & 4.5 & 9.4 & 2.8 & 2.5 & 18.7 & 4.6 & 4.4 & 13.2 & 3.0 & 2.8 \\
star11 & 19.4,18.9 & 20.2 & 6.1 & 4.4 & 9.6 & 2.7 & 2.4 & 14.4 & 4.7 & 4.3 & 13.9 & 4.4 & 2.8 \\
\enddata
\end{deluxetable}

\clearpage

\begin{deluxetable}{cccccccc}
\tabletypesize{\footnotesize}
\tablewidth{0pt}
\rotate
\tablecaption{Parameters of Double-Sinusoid Fits to [4.5] Light Curves}
\tablehead{ 
\colhead{Epoch UTC}  & \multicolumn{2}{c}{Semi-Amplitude \%}  & \multicolumn{2}{c}{Period (hours)} &  \multicolumn{2}{c}{Phase (radians)} & Constant \\
2013      & \colhead{$A_1$} & \colhead{$A_2$} & \colhead{$P_1$} & \colhead{$P_2$\tablenotemark{a}} &
 \colhead{$\phi_1$} &   \colhead{$\phi_2$} & C \\
}
\startdata
June 30  & 0.011 $\pm$ 0.001 & 0.003 $\pm$ 0.001 & 5.8 $\pm$ 0.1 & 2.9  & 2.1 $\pm$ 0.4 & 6.2 $\pm$ 0.9 & 0.9977 $\pm$ 0.0007 \\
October 30  & 0.0072 $\pm$ 0.0009 & 0.0066 $\pm$ 0.0009 & 6.13 $\pm$ 0.08 & 3.07  & 3.9 $\pm$ 0.3 & 0.7 $\pm$ 0.5 & 0.9984 $\pm$ 0.0006 \\
\enddata
\tablenotetext{a}{The period of the second sinusoid is forced to be half the first in our fitting procedure, see text. The light curve is described by:
$$F(t) = A_1\sin\left(\frac{2\pi}{P}t+\phi_1\right) + A_2\sin\left(2\frac{2\pi}{P}t+\phi_2\right) + C$$
where $A$ is semi-amplitude in \%, $P$ is period in hours, $t$ is time in hours, $\phi$ is phase in radians and $C$ is a constant. The parameter and uncertainty values given in the Table correspond to the  16th, 50th, and 84th percentiles of the marginalized distributions.
}
\end{deluxetable}

\clearpage

\begin{deluxetable}{lcccccc}
\tabletypesize{\footnotesize}
\tablewidth{0pt}
\tablecaption{T Dwarfs with Both Near- and Mid-Infrared Variability Studies}
\tablehead{ 
\colhead{Name}  & \colhead{Spectral} & \multicolumn{2}{c}{Near-Infrared}  & \multicolumn{2}{c}{Mid-Infrared} &  \colhead{References} \\
   & \colhead{Type} & \colhead{Amplitude} & \colhead{Period} & \colhead{Amplitude\tablenotemark{a}} & \colhead{Period} &  \\
   & & \% & hour & \% & hour & \\
}   
\startdata
SDSS J015141.69$+$124429.6 & T1 & $\leq 1.1$\tablenotemark{b} &   \nodata &   
$< 0.6$  &   \nodata & Rad14, Met15 \\
2MASS J21392676+0220226 & T1.5 &  8--26 & 7.72 &  $\sim$11 & 7.61 & Ra12, Ya16 \\
SDSSp J125453.90$-$012247.4 & T2 &  $\leq 2.1$\tablenotemark{c} &   \nodata &    $< 0.3$ &   \nodata & Ra14, Me15 \\
SIMP J013656.5$+$093347.3 & T2.5 & 6 & 2.4 & $\sim$6 & 2.41 & Ar09, Ra14, Ya16 \\
2MASS J13243553$+$6358281\tablenotemark{d} & T2.5 &  17 & 13.2 & 3 & 13 & He15, Me15, Ya16 \\
2MASSI J2254188$+$312349 & T4 &   $\leq 0.8$\tablenotemark{e} &   \nodata &   $< 0.5$ &   \nodata & Ra14,  Me15 \\
2MASS J22282889$-$4310262 & T6 & 1.6 & 1.42 & 4.6 & 1.41 & Ra14, Me15 \\
2MASS J00501994$-$3322402 & T7 &  10.8\tablenotemark{f} & \nodata & 1.1 & 1.55 &
Wi14, Me15 \\
Ross 458C & T8 & $< 2.1$ & \nodata & $< 1.4$ & \nodata &
Ra15, Me15 \\
%
%
%
\enddata
\tablenotetext{a}{Metchev et al. (2015) give variability amplitudes for both [3.6] and [4.5]; the larger of the two is listed in the table.}
\tablenotetext{b}{Enoch et al. (2003) measured  45\% variability in $K_s$ for one epoch, but observations at other times have found the object to not vary.}
\tablenotetext{c}{Goldman et al. (2008) report possible variations seen in spectra at $\lambda \sim 1.1 ~\mu$m and 1.6~$\mu$m.} 
\tablenotetext{d}{Burgasser et al. 2010 suggest that 2MASS J13243553$+$6358281 may be a close L dwarf and T dwarf binary.} 
\tablenotetext{e}{Enoch et al. (2003) measured  56\% variability in $K_s$ for one epoch, but observations at other times have found the object to not vary.}
\tablenotetext{f}{Re-analysis of the Wi14 near-infrared data by Radigan (2014) determined a variability amplitude of $< 0.7$\%.}
\tablecomments{Amplitudes are peak-to-peak. References are Artigau et al. 2009; Heinze et al 2015; Metchev et al. 2015; Radigan et al. 2012, 2014; Rajan et al. 2015; Wilson et al. 2014; Yang et al. 2016.}
\end{deluxetable}


\begin{thebibliography}{}
\bibitem[Artigau et al. (2009)]{}Artigau, E., Bouchard, S., Doyon, R., \& Lafreniere, D. \ 2009, \apj, 701, 1534
\bibitem[Beichman et al. (2014)]{} Beichman, C. et al. \ 2014, \apj, 783, 68
\bibitem[Burgasser et al. (2010)]{}Burgasser, A. et al. \ 2010, \apj, 710, 1142
\bibitem[Burrows et al. (1997)]{}       
Burrows, A., Marley, M., Hubbard, W. B., Lunine, J. I., Guillot, T., Saumon, D., Freedman, R., Sudarsky, D. \& Sharp, C. \ 1997, \apj, 491, 856
\bibitem[Burrows, Sudarsky \& Lunine (2003]{}Burrows, A., Sudarsky, D. \& Lunine, J. I. \ 2003, \apj, 596, 587
\bibitem[Crossfield (2014)]{}Crossfield, I. \ 2014, \aap, 566, A130
\bibitem[Cushing et al. (2011)]{ys} Cushing, M. C. et al. \ 2011, \apj, 743, 50
\bibitem[Cushing et al. (2014)]]{}Cushing, M. C., Kirkpatrick, J. D., Gelino, C. R., Mace, G. N., Skrutskie, M. F. \& Gould, A. \ 2014, \aj, 147, 113
\bibitem[Cushing et al. (2016)]{}Cushing, M. C. et al. \ 2016, \apj, in press (Cu16)
\bibitem[Enoch et al. (2003)]{}Enoch, M. L., Brown, M. E. \& Burgasser, A. J., \ 2003, \aj, 126, 1006
\bibitem[Fazio et al. (2004)]{}Fazio, G. G., et al. \ 2004, \apjs, 154, 10
\bibitem[Gelino \& Marley (2000)]{} Gelino. C. \& Marley, M. S. \ 2000,
From Giant Planets to Cool Stars, ASP Conference Series, Vol. 212. Edited by Caitlin A. Griffith and Mark S. Marley.
\bibitem[Goldman et al. (2008)]{}Goldman, B. et al., \ 2008, \aap, 487, 277
\bibitem[Heinze et al. (2015)]{}Heinze, A. N., Metchev, S. \& Kellogg, K. \ 2015, \apj, 801, 104 
\bibitem[Herbst (1989)]{}Herbst, W., \ 1989, \aj, 98, 2268
\bibitem[Hodapp et al. (2003)]{NIRI} Hodapp, K. W. et al.\ 2003, \pasp, 115, 1388
\bibitem[Hogg, Bovy \& Lang (2010)]{}Hogg D. W., Bovy J. \& Lang D. 2010 arXiv:1008.4686
\bibitem[Kirkpatrick et al. (2012)]{} Kirkpatrick, J. D. et al. \ 2012, \apj, 753, 156
\bibitem[Leggett et al. (2013)]{} Leggett, S. K. et al. \ 2013, \apj, 763, 130
\bibitem[Leggett et al. (2016)]{} Leggett, S. K. et al. \ 2016, \apj, in press
\bibitem[Liu et al. (2012)]{}Liu, M. C., Dupuy, T. J., Bowler, B. P., Leggett, S. K. \& Best, W. M. J. \ 2012, \apj, 758, 57
\bibitem[Luhman (2014)]{}Luhman, K. L.  \ 2014, \apj, 786, L18
\bibitem[Luhman, Burgasser \& Bochanski (2011)]{} Luhman, K. L., Burgasser, A. J. \& Bochanski, J. J. \ 2011,  \apj, 730, L9 
\bibitem[Marley \& Sengupta (2011)]{}Marley, M. S. \& Sengupta, S. \ 2011, \mnras, 417, 2874
\bibitem[Metchev et al. (2015)]{}Metchev S. A., et al. \ 2015, \apj, 799, 154
\bibitem[Morley et al. (2012)]{}Morley, C. V., Fortney, J. J., Marley, M. S., Visscher, C., Saumon, D. \& Leggett, S. K. \ 2012, \apj, 756, 172
\bibitem[Morley et al. (2014)]{}Morley, C. V., Marley, M. S.,Fortney, J. J., Lupu, R., Saumon, D., Greene, T. \& Lodders, K. \ 2014,  \apj, 787, 78
\bibitem[Pinfield et al. (2014)]{}Pinfield, D. J. et al. 2014, \mnras, 444, 1931
\bibitem[Radigan et al. (2012)]{}Radigan, J., Jayawardhana, R., Lafreniere, D., Artigau, E., Marley, M. S. \& Saumon, S. \ 2012, \apj, 750, 105
\bibitem[Radigan(2014)]{}Radigan, J., \ 2014,  \apj, 797, 120
\bibitem[Radigan et al. (2014)]{}Radigan, J., Lafreniere, D., Jayawardhana, R. \& Artigau, E., \ 2014, \apj, 793, 75
\bibitem[Rajan et al. (2015)]{}Rajan, A. , et al. \ 2015, \mnras, 448, 3775
\bibitem[Robinson \& Marley (2014)]{}Robinson, T. D. \& Marley, M. S. \ 2014, \apj, 785
\bibitem[Saumon \& Marley (2008)]{}Saumon, D. \& Marley, M. S. \ 2008, \apj, 689, 1327
\bibitem[Schneider et al. (2015)]{}Schneider A. C. et al. \ 2015, \apj, 804, 92
\bibitem[Showman \& Kaspi (2013)]{}Showman, A. P. \& Kaspi, Y. \ 2013, \apj, 776, 85
\bibitem[Simon et al. (2016)]{}Simon, A. A. et al. \ 2016, \apj, 817, 162
\bibitem[Sromovsky et al. (2012)]{}Sromovsky L. A., et al. \ 2012, \icarus, 220, 6
\bibitem[Tackett, Herbst \& Williams (2003)]{}Tackett, S., Herbst, W. \& Williams, E. \ 2003, \aj, 126, 348
\bibitem[Tinney et al. (2012)]{}Tinney C. G., Faherty, J. K., Kirkpatrick, J. D., Wright, E. L., Gelino, C. R.; Cushing, M. C.; Griffith, R. L. \& Salter, G. \ 2012, \apj, 759, 60
\bibitem[Tremblin et al. (2015)]{}Tremblin, P., Amundsen, D. S., Mourier, P., Baraffe, I., Chabrier, G., Drummond, B., Homeier, D. \& Venot, O. \ 2015, \apj, 804, L17 
\bibitem[Wilson et al. (2014)]{}Wilson, P. A., Rajan, A. \& Patience, J., \ 2014, \aap, 566, A111 
\bibitem[Yang et al. (2016)]{}Yang, H. et al. \ 2016,  2016arXiv160502708Y
\bibitem[Zhou et al. (2016)]{}Zhou, Y., Apai, D., Schneider, G. H., Marley, M. S. \& Showman, A. P. \ 2016, \apj, 818, 176

\end{thebibliography}
\end{document}